\PassOptionsToPackage{dvipsnames}{xcolor}
\documentclass[a4paper,11pt,final]{article}
\usepackage{import}
\usepackage{Preamble}

\title{
Scrambling or Stalling: Angular Momentum Barriers to Chaos in Holographic CFTs
}

\author[a,b]{Juan Hernandez}
\author[a]{and Andrew Rolph}

\affiliation[a]{Vrije Universiteit Brussel (VUB) and The International Solvay Institutes, Pleinlaan 2, B-1050 Brussels, Belgium}

\affiliation[b]{School of Mathematics and Hamilton Mathematics Institute, Trinity College, Dublin 2,
Ireland}

\emailAdd{hernanju@tcd.ie}
\emailAdd{andrew.d.rolph@gmail.com}

\abstract{
Scrambling is a diagnostic of quantum chaos in strongly coupled systems, and plays a central role in the holographic description of black hole dynamics. We study scrambling in high-temperature holographic CFTs, with an emphasis on perturbations dual to particles on infalling and bound trajectories in the bulk description. For BTZ and AdS-Schwarzschild geometries, we derive an analytic expression relating the difference in scrambling times to the particles' kinematics. We match this to a 2d CFT computation by constructing the smeared operator that creates the bulk particle with the desired kinematics and calculating the out-of-time-ordered correlator (OTOC). For higher-dimensional holographic CFTs, the scrambling slows and eventually ceases when the dual bulk particle has insufficient energy to overcome the angular momentum barrier. 
}

\makeatletter
\gdef\@fpheader{}
\makeatother
\begin{document}
\nolinenumbers
\maketitle
\flushbottom

\section{Introduction}

Is quantum gravity a chaotic theory?
There are several diagnostic signatures of quantum chaos, and a variety of gravitational systems exhibit those signatures.
For example:
(1) black holes are the fastest scramblers, with a scrambling time of order $\log S$~\cite{Hayden:2007cs, Sekino:2008he, Lashkari:2011yi},
(2) the energy spectrum of (JT) gravity displays eigenvalue repulsion~\cite{Cotler:2016fpe, DAlessio:2015qtq},
and (3) there is strong sensitivity to initial conditions: small perturbations can lead to shockwaves near a black hole horizon~\cite{DRAY1985173, aichelburg1971gravitational, Sfetsos:1994xa}, with the strength of the shockwave growing as $e^{\frac{2\pi}{\beta}t}$~\cite{Shenker:2013pqa}.

In holography, AdS black holes are dual to thermal states in the boundary conformal field theory (CFT), and this holographic CFT is strongly coupled, so it is natural to expect generic perturbations to rapidly thermalise and scramble. This is dual to the ringdown of quasinormal modes and perturbations falling into the black hole. 

Yet not all perturbations in a black hole background exhibit chaotic dynamics. 
In particular, localised particle excitations whose angular momentum is above a critical $J_{\text{crit.}}$ will not come closer than the photon sphere and so will not thermalise or scramble. In the bulk, there is a straightforward understanding: there are both near-horizon quasinormal modes and long-lived approximate normal modes trapped outside of the photon sphere~\cite{Festuccia:2008zx}. 
But, from the perspective of the boundary CFT, this behaviour is surprising. The kinematics of the bulk particle is controlled by how the boundary operator is smeared, and it is not a priori clear, from the CFT perspective, that a certain continuous deformation of the smearing kernel would lead to a rapid cessation of
scrambling behaviour
at $J_{\text{crit.}}$, and for the boundary operator to dynamically oscillate in size~\cite{Susskind:2018tei, Haehl:2021emt}. 
Absent the dual holographic description, this non-ergodic behaviour would be surprising in a strongly coupled thermal CFT.

In this paper, we explore the difference in chaotic dynamics in holographic CFTs for different perturbations. On the bulk side, we compare how perturbations following different trajectories scramble in AdS$_{d+1}$ black hole backgrounds, and we match the $d=2$ result to a thermal CFT$_2$ calculation similar to that of~\cite{Roberts:2014ifa}. 
To probe the chaotic behaviour of these perturbations, we use the four-point out-of-time-ordered correlator (OTOC) between a pair of operators:
\begin{equation}\label{eq: OTOC}
   \langle W(\tW) V(0) W(\tW) V(0) \rangle_\beta \,.
\end{equation}
where $\langle (\dots)\rangle_\beta = Z_\beta^{-1} \Tr (e^{-\beta H} (\dots))$, and $Z_\beta = \Tr(e^{-\beta H})$. 

The OTOC has been extensively used in the study of quantum chaos, both from the field theory~\cite{Haake:2010fgh, Roberts:2014ifa, Rozenbaum:2016mmv, Hashimoto:2017oit, Nahum:2017yvy, Hampapura:2018otw, Demulder:2025uda}
and gravitational perspective~\cite{Roberts:2014isa, Shenker:2014cwa, Craps:2020ahu}. 
To understand the OTOC's relation to chaos, first note that, in classical systems, the sensitivity to initial conditions, the butterfly effect, is quantified by the Poisson bracket $\{x(t),p(0)\} = \frac{\partial x(t)}{\partial x(0)}$, which grows exponentially in time for chaotic systems. In quantum systems, the analogue to the Poisson bracket is the squared-commutator, which is closely related to, and inherits its exponential growth from, the OTOC:%
\footnote{We take $V$ and $W$ to be Hermitian. Then $-[W(\tW),V(0)]^2$ is positive semi-definite. Positivity needs the minus sign because $[W, V]$ is anti-Hermitian. Also, $\langle W(\tW) W(\tW) V(0) V(0) \rangle_\beta$ approximates to the $\tW$-independent $\langle W(\tW) W(\tW) \rangle_\beta \langle V(0) V(0) \rangle_\beta$ for $\tW \gg \beta$. }
\begin{equation}\label{eq: com squared}
    - \langle [W(\tW),V(0)]^2\rangle_\beta = 
 2 \langle W(\tW) W(\tW) V(0) V(0) \rangle_\beta - 2 \, {\rm Re}\, G(\tW) \,.
\end{equation}

The OTOC also quantifies scrambling and operator growth. For generic, few-body, operators $V$ and $W$ that initially commute, $[V(0),W(0)] = 0$,
the squared-correlator is initially zero. But if $W$ is moved further to the past ($\tW \leq0$) then, for a Hamiltonian with local interactions, 
the time-evolved operator $W(\tW)$ has more time to grow. The scrambling time $t_*$ is the value of $-\tW$ at which point $W(\tW)$ has grown enough that it no longer commutes with generic operators
$V(0)$, leading to a non-zero and growing squared-commutator. In a highly chaotic theory, this gives an exponentially decaying OTOC: 
\bne\label{eq: OTOC expansion} \frac{\langle W(\tW) V(0) W(\tW) V(0) \rangle_\beta}{\langle W(\tW) W(\tW)\rangle_\beta \langle V(0) V(0) \rangle_\beta} \approx 1 - \frac{a}{N_{\rm{eff.}}}\, e^{-\lambda_L \tW},\qquad \beta \ll -\tW \ll t_* \ene

If $N_{\rm{eff.}}$ is the effective number of degrees of freedom, then the scrambling time $t_*$ scales as $ \lambda_L^{-1} \log (N_{\rm{eff.}})$. 
Note also that the OTOC equals the overlap between the two (unnormalised) states $W V \ket{\text{TFD}}$ and $V W\ket{\text{TFD}}$, and it is the failure of $V$ and $W$ to commute that causes this overlap to decrease. 
One of the main goals of the present work is to quantify how scrambling depends on certain details of the initial perturbation, with particular emphasis on perturbations dual to bulk particles which follow classical orbits around the black hole geometry. 

In Sec.~\ref{sec: bulk}, we start the investigation from the bulk side, and consider particles released from the boundary of non-rotating BTZ and AdS$_{d+1}$-Schwarzschild geometries with different energy and angular momenta. For AdS-Schwarzschild, particles with angular momenta above a critical value $J_{\text{crit.}}$ do not fall into the black hole, and instead follow a radially-oscillating bound orbit that periodically returns to the boundary. Correspondingly, the perturbation fails to scramble, and the squared-commutator remains small. For particles that do fall in, we determine the dependence of the particles' scrambling time on their energy and momenta from the resulting shockwave geometries. 

For example, for global BTZ, the difference in scrambling times for two particles is
\bne t_*^{(2)} - t_*^{(1)} = \frac{1}{r_h} \log \left (  \sqrt{\frac{E_1^2-J_1^2}{E_2^2-J_2^2}} \frac{\cosh(r_h(\pi - \tilde{\varphi}^{(1)}))}{\cosh(r_h(\pi - \tilde{\varphi}^{(2)}))} \right ) \ene
where $r_h$ is the horizon radius, and
\bne \tilde{\varphi}:= \left( \phiV -\phiW - \frac{1}{r_h} \arctanh \left(\frac{J}{E}\right ) \right) \mod{2\pi},
\label{eq:diffst}
\ene
with $\phiV$ and $\phiW$ the operator insertion positions on the boundary circle.

We derive this from a bulk calculation, and generalise to higher-dimensional AdS black holes, where there is a critical $J_{\text{crit.}}$ above which particles no longer fall into the black hole. For the particles that do fall in, as the angular momentum approaches $J_{\rm crit.}$, the delay in scrambling time diverges, 
interpolating between the scrambling and non-scrambling regimes.

In Sec.~\ref{sec: particles}, we determine the boundary operators that create an approximately classical bulk particle with a given energy and boundary-parallel momentum. The local operator is smeared over a kernel $K$, $W_K = \int K W$, and the kernel is found using bulk Gaussian wavepacket solutions and the extrapolate dictionary. The resulting kernel for a particle without momentum is given in~\eqref{eq:smea1}. This derivation is based on~\cite{Polchinski:1999ry}, but see also~\cite{Fitzpatrick:2011jn, Papadodimas:2012aq, Leichenauer:2013kaa, Terashima:2021klf, Terashima:2023mcr, Tanahashi:2025fqi} on bulk particle wavepackets, and~\cite{Caron-Huot:2025hmk} on the boundary kernel for Gaussian wavepackets.
For particles with non-zero momentum, the kernel can be found by an appropriate translation and boost such that the insertion point remains unchanged, see eqs.~\eqref{eq:bstsm} and \eqref{eq:boske}. 

In Sec.~\ref{sec: CFT}, we reproduce the planar BTZ result from a CFT$_2$ calculation, building on~\cite{Roberts:2014ifa}. We compute the OTOC between the smeared operator $W_K$ and a local operator $V$, with $K$ the kernel derived in Sec.~\ref{sec: particles} to give the dual $W$-particle a particular energy and boundary-parallel rapidity. 
Compared to the CFT$_2$ OTOC of local operators~\cite{Roberts:2014ifa}, the Lyapunov exponent  $\lambda_L = \frac{2\pi}{\beta}$ and butterfly velocity $v_b=1$ are unaffected, but the $O(c^0)$ part of the scrambling time \textit{is} sensitive to the smearing kernel. For example, comparing two excitations with the same energy, but one with rapidity $\eta$, the difference in scrambling times is
\begin{equation}\label{eq: Delta t result}
    t^{(2)}_* - t^{(1)}_* =  \left |\xW+ \frac{\beta}{2\pi}\eta \right| - |\xW| + \frac{\beta}{2\pi} \log(\cosh(\eta))\,.
\end{equation}
The first two terms show that the butterfly cone has been shifted by $\frac{\beta}{2\pi} \eta$, which corresponds to the $x$-distance the perturbation travels before reaching the black hole horizon.
The last contribution to~\eqref{eq: Delta t result} is a delay in the scrambling, dual to the time needed for the bulk particle to reach a given blue-shifted energy, because the perturbations start at different radii. 
Both contributions to the change in scrambling time increase linearly for large $\eta$.  

Perhaps the most intriguing aspect of this work is the prediction for operator dynamics in higher-dimensional CFTs; in particular, the sharp transition in the OTOC for perturbations with $J > J_{\text{crit.}}$, and the oscillating operator size. Unfortunately, we are not able to understand and confirm these predictions with a direct CFT calculation because of the difficulty of calculating OTOCs in higher-dimensional CFTs. 
This, and other directions for future work, will be discussed in Sec.~\ref{sec:discussion}. 

\section{Bulk calculation of the scrambling time}\label{sec: bulk}
\begin{figure}
    \centering
    \includegraphics[width=0.9\linewidth]{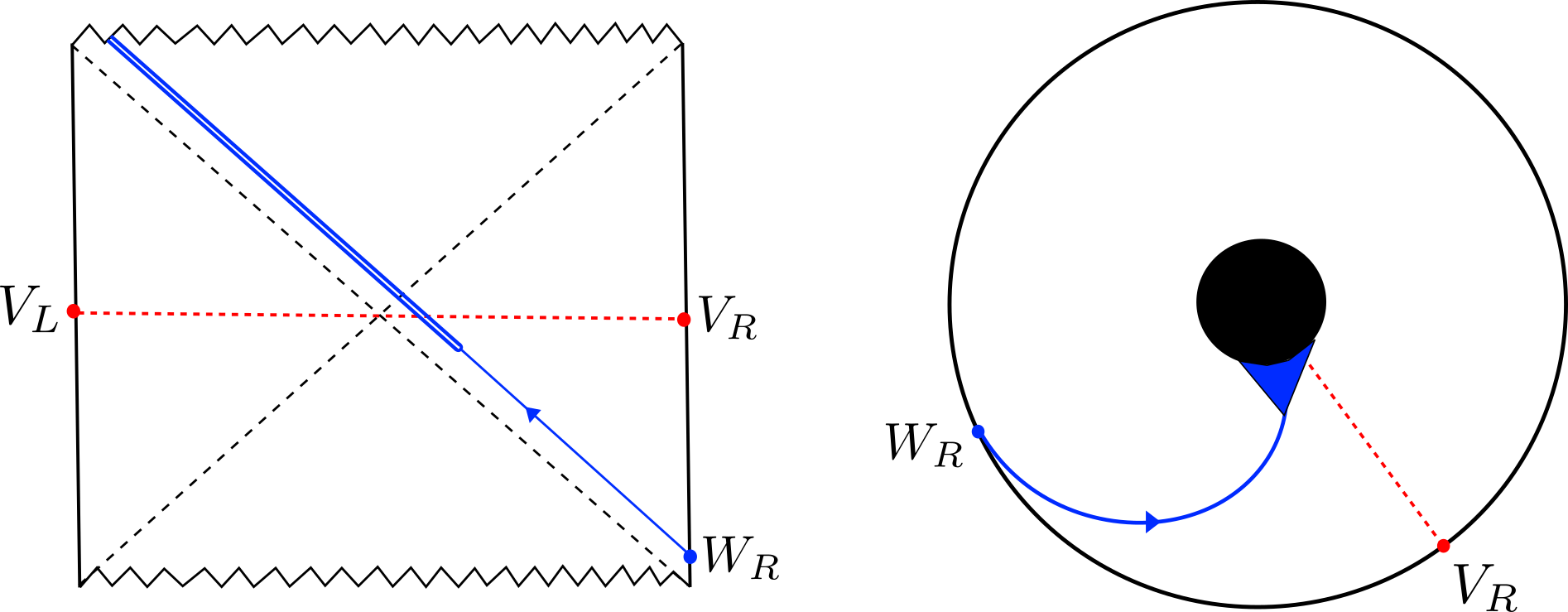}
    \caption{We release a $W$-particle with some energy and angular momentum from near the AdS$_R$ boundary. This leads to a shockwave backreaction near the black hole horizon, and scrambling of the $W$-perturbation, as quantified by the $\bra{\text{TFD}}W_R V_L V_R W_R \ket{\text{TFD}}$ correlator. The scrambling time depends on the energy and momentum of the $W$-particle and the relative positions of the operators. The diagram on the right is the $t=0$ slice of the left-hand bulk's geometry, with the blue triangle representing the growing near-horizon shockwave.}
    \label{fig:BulkOTOC}
\end{figure}

Consider a BTZ or AdS-Schwarzschild black hole and a particle released near the asymptotic boundary. If the particle falls into the black hole, then there is a near-horizon blueshift of energy, which leads to a shockwave and scrambling as measured by a boundary probe operator. In this section, we will calculate the dependence of the scrambling time on the conserved energy and angular or linear momentum of the particle, and the relative position of the probe operator. To be more precise, we will calculate the difference in scrambling time for two particles with different energies and angular momenta. See Fig.~\ref{fig:BulkOTOC} for an illustration of the setup. 

\subsection{Geometry and geodesics}
We start by giving the formulas for particle motion in black hole geometries that we will need for the rest of the section. We will start with general static and spherically symmetric geometries, then specialise to BTZ and AdS-Schwarzschild.

The metric of a static and spherically symmetric $(d+1)$-dim Schwarzschild geometry in global coordinates is 
\bne ds^2 = - f(r) dt^2 + \frac{dr^2}{f(r)} + r^2 (d\phi^2 + \sin^2 \phi \, d\Omega^2_{d-2}) .
\label{eq:metbh}
\ene
We assume that there is a single horizon, $f(r_h) =0$, with $r_h$ the horizon radius. The surface gravity at the horizon $\kappa$ and the horizon temperature are related by $\kappa = \frac{1}{2} f'(r_h) = 2\pi T$, and we assume that $T$ is non-zero.

We will consider only geodesics tracing curves in the $(r,\phi)$ plane, which is without loss of generality because of the rotational symmetry.
The Lagrangian for geodesic motion is $L = \frac{1}{2} g_{\mu\nu} \dot x^\mu \dot x^\nu$, 
for which the conserved momenta in the geometry~\eqref{eq:metbh} are 
$ p_t = -f(r) \dot{t}$ and $J = p_\phi =  r^2 \dot \phi$. 
The general formula for the energy of a particle is
$ E= -g_{\mu\nu} \xi^\mu p^\nu$, where $\xi$ is a timelike Killing vector field; for the geometry~\eqref{eq:metbh}, if we choose $\xi = \del_t$, then $E = f(r) p^t = f(r) \dot{t}$. 

We take our particles to have an energy $E$ much larger than the AdS and thermal energy scales. In geometries that are asymptotically AdS, they are released from high up in the AdS radial potential, at $r \approx E$ (we take $l_{AdS} =1$), and, even if they are massive, they become relativistic from rest on timescales $\Delta t \approx E^{-1} \ll 1$. So, we will approximate our particle trajectories as null rays.

For a null geodesic, starting from $r=\infty$ at $\tW$, the time taken to reach a given radius $r$ is 
 \bne   t(r) - \tW
=  \int_r^\infty dr' \frac{1}{f(r') \sqrt{1- \frac{J^2}{E^2}\frac{f(r')}{r'^2}}} .
 \label{eq:unitr}\ene
Due to the gravitational redshift, this time difference diverges logarithmically as the geodesic approaches the horizon:
\bne t(r) -\tW = \frac{1}{2\kappa}\left[- \log (r-r_h)+ \log(A^2) + O(r-r_h)\right] , \qquad r \to r_h. \label{eq:tlargr} \ene
The constant $A$ is $f$-dependent, and can be evaluated analytically for BTZ, but not for higher-dimensional AdS-Schwarzschild black holes, except in certain limits, such as large $d$. 
Inverting \eqref{eq:tlargr}, we get%
\footnote{We use $\sim$ in a precise way, to denote asymptotic equivalence: $f(x) \sim g(x)$ as $x\to \infty$ iff $\lim_{x\to\infty}\frac{f(x)}{g(x)} = 1$.}
\bne r(t) -r_h  \sim A^2 e^{-2\kappa (t-\tW)} , \qquad (t-\tW) \to \infty .\ene

We will also need the geometry~\eqref{eq:metbh} in Kruskal coordinates. The coordinate transformation we will use is $U := - e^{- \kappa (t-r_*)}$ and $V:= e^{\kappa (t+r_*)}$, 
and where the tortoise coordinate is
\bne \begin{split} 
r_* (r) := - \int_r^\infty \frac{dr'}{f(r')} 
\end{split} \ene
With this convention for the tortoise coordinate's additive constant, we have $r_* (\infty) = 0$. Then, as the geodesic approaches the horizon, $r \to r_h$, 
\bne \begin{split} 
r_* (r) &= \frac{1}{2\kappa}\log\left(B^2(r-r_h)\right) + O((r-r_h)^1) \\
&= -(t-\tW) + \frac{\log (AB)}{\kappa} + O((\tW -t)^{-1}).
\label{eq:nhtco}
\end{split}\ene
$B$ is another undetermined $f$-dependent constant, but, unlike $A$, it does not depend on $E$ or $J$, and it will drop out when we calculate the difference in scrambling times. In terms of the Kruskal coordinate $U$, this gives us 
\bne \begin{split} U(t) &= - e^{\kappa (r_* (t) -t)}, \\
&\sim -A B e^{-2\kappa t} e^{\kappa \tW},\qquad \text {as } (t-\tW)\to\infty .
\label{eq:Ucoord} 
\end{split}\ene
So, for fixed $t$, the particle approaches the $U = 0$ outer future horizon exponentially fast as $\tW \to -\infty$. This is what will lead to the exponential growth of the particle's $T_{UU}$, which creates a shockwave.

\subsubsection{BTZ}
For a BTZ black hole, 
$f (r) = (r^2 - r_h^2)$, and
the ADM mass and temperature are related to the horizon radius by $r_h^2 = 8 G_N M$ and $\kappa = r_h = 2\pi T$.

The Lagrangian for a geodesic can be written as $\dot{r}^2 + V(r) = E^2$, where the radial potential is
\bne V(r) = a (r^2 - r_h^2) + J^2 \left(1- \frac{r_h^2}{r^2}\right) \ene
The constant $a$ is zero for null geodesics, and one for timelike geodesics. 
Physical particles have $|J| < E$.
This radial potential increases monotonically for both null and timelike geodesics, so, unlike AdS-Schwarzschild black holes, all massless and massive particles fall into the BTZ black hole. The same is also true of rotating and quantum BTZ black holes.

For BTZ, we can evaluate~\eqref{eq:unitr} and calculate $t(r)$ exactly. The near-horizon expansion is
\bne t(r) - \tW = -\frac{1}{2r_h} \log (r-r_h) - \frac{1}{2r_h} \log \left (\frac{E^2 - J^2}{2E^2 r_h} \right)+ O(r-r_h). \label{eq:trgir} \ene
This shows that increasing $J$ increases the time to reach a given radius, diverging as $|J| \to E$, as expected. Eq.~\eqref{eq:trgir} also gives us 
\bne A^2 = \frac{2r_h}{1-J^2/E^2} \ene 
and
\bne r(t) - r_h=  \frac{2r_h}{1-J^2/E^2} e^{-2r_h (t-\tW)} + O(e^{-4r_h (t-\tW)}) .\ene

We will need the angle at which a null geodesic from $r= \infty$ at $\phi = \phiW$ reaches the BTZ horizon. Solving $d\phi/dr = \dot{\phi}/\dot{r}$ gives
\bne \phi_h := \lim_{r \to r_h} \phi(r) = \left (\phiW + \frac{1}{r_h} \arctanh \left( \frac{J}{E}\right )\right ) \mod{2\pi} \label{eq:phihd} .\ene

For BTZ, the tortoise coordinate is 
\bne r_* (r) = \frac{1}{2r_h}\log \left( \frac{r-r_h}{r+r_h} \right), \ene
which gives us $B^2 = \frac{1}{2r_h}$, 
and the metric in Kruskal coordinates, which is
\bne ds^2 = -\frac{4}{(1+UV)^2} dU dV + r_h^2 \left (\frac{1-UV}{1+UV} \right)^2 d\phi^2. \ene

\subsubsection{AdS-Schwarzschild}

The emblackening factor $f$ for AdS$_{d+1}$-Schwarzschild is
\bne f(r) = 1+r^2 - \frac{\mu}{r^{d-2}} .\ene
The mass parameter $\mu$ and the surface gravity $\kappa$ are related to the horizon radius $r_h$: $\mu = r_h^{d-2} (1+r_h^2)$, and $\kappa = \frac{d-2}{2r_h} + \frac{d}{2} r_h$. In contrast to BTZ, we cannot calculate $r_* (r)$, $\phi_h$ or $A$ analytically for arbitrary $d$, though all can be determined numerically, and perturbatively in certain limits, such as small $\mu$ or large $d$. 

Unlike the BTZ black hole, particles can avoid falling into the AdS-Schwarzschild black hole. From the radial potential, 
\begin{equation}
    V(r) = \left(a+\frac{J^2}{r^2}\right) f(r)\end{equation}
we can determine the critical angular momentum $J_{\text{crit.}}$ below which particles will fall in. For a null geodesic ($a=0$),
\bne \frac{J_{\text{crit.}}}{E} = \frac{1}{\sqrt{\frac{d-2}{d} \frac{1}{r_{\text{ph}}^2}+1}} \label{eq:Jcrit} \ene
where $r_{\text{ph.}}$ is the photon sphere radius
\bne r_{\text{ph.}} = \left(\frac{d\mu}{2}\right)^{\frac{1}{d-2}} .\ene

For trajectories with $J^2/J_{\rm crit.}^2 = 1-\epsilon$, there is an additional logarithmic divergence in~\eqref{eq:unitr}, from the time it takes the particle to slowly roll over the angular momentum barrier: 
\begin{equation}
\begin{aligned}\label{eq: delta t div}
    t(r)-\tW =  \alpha \log(\epsilon^{-1})+O(\epsilon^{0}) \,,\qquad \alpha = \frac{\sqrt{2}E}{f(r_{\rm ph})\sqrt{-V''(r_{\rm ph})}}.
\end{aligned}
\end{equation}
This immediately implies a $\log(\epsilon^{-1})$ divergence in the scrambling time as $J \to J_{\rm crit.}$. In section~\ref{sec: scrambling time}, this divergence is captured by the fact that the $A^{(E,J)}$ coefficient diverges for $J \to J_{\rm crit.}$.

For large AdS black holes, $\mu \gg 1$, $J_{\text{crit.}}/E \approx 1$, while for small AdS black holes, $\mu \ll 1$,
\bne \frac{J_{\text{crit.}}}{E} \approx \sqrt{\frac{d}{d-2}} \left( \frac{d}{2} \right )^{\frac{1}{d-2}} r_h .\ene
This shows that, for small AdS black holes, we do not need $J_{\text{crit.}}/E$ to be close to its maximal value of one for the particle to miss the black hole. Also, we have 
\bne \left( \frac{r_{\text{ph.}}}{r_h} \right )^{d-2} = \frac{d}{2} (1+ r_h^2) > 1, \ene 
and so relativistic particles with $J > J_{\text{crit.}}$
do not come close to the horizon. Thus, there is no significant blueshift of energies or backreaction, and the boundary operator will detect no $O(G_N^0)$ scrambling of the TFD state. 
This predicts qualitatively different behaviour in the boundary OTOC, such as the divergence of
the scrambling time as $J\to J_{\rm crit.}$, when the perturbing $W$ operator approaches and exceeds $J = J_{\text{crit.}}$.

\subsection{Particle stress tensor}
For now, let us assume that $J < J_{\text{crit.}}$ so that the boundary-released particle falls into the black hole.
To calculate the backreaction,
we will need the particle stress tensor, which is%
\footnote{The stress tensor~\eqref{eq:einbeinT} is valid for massive and massless particles, but to get the canonical form of the massive particle stress tensor, one makes the gauge choice $e^{-1} = m$, which reduces the equation of motion~\eqref{eq:eom} to $\dot{x}^2 = -1$; this gauge choice is equivalent to choosing a normalisation for the massive particle's timelike velocity.}
\bne \label{eq:einbeinT} T^{\mu\nu}(y) =  \int ds \frac{1}{e(s)} \frac{\delta^d (y - x(s))}{\sqrt{-g}} \dot{x}^\mu \dot{x}^\nu .\ene
This comes from the action
\bne S = \frac{1}{2} \int ds (e^{-1} \dot{x}^2 -e m^2), \ene
which has an equation of motion
\bne \label{eq:eom} \dot{x}^2 + e^2 m^2 = 0 \ene
and momenta
\bne \label{eq:pp} p^\mu = e^{-1}\dot{x}^\mu .\ene

We will show that the $T_{UU}$ component of the stress tensor diverges as the particle approaches the outer future horizon.
From~\eqref{eq:einbeinT} and~\eqref{eq:pp}, suppressing all angular directions except for $\phi$,
\bne T_{UU} (U,V,\phi) = - \delta (\phi - \phi(V)) \delta (U-U(V)) \frac{g_{UV}}{\sqrt{g_{\phi\phi}}} p^V . \label{eq:TUUeq} \ene
Next, we determine $p^V$ by eliminating $p^U$ from the pair of equations $g_{\mu\nu}p^\mu p^\nu = 0$ and $E = \kappa (U p_U - V p_V)$. Taking the $U \to 0^-$ limit of the result shows that $p^V$ diverges as the particle approaches the $U = 0$ horizon:
\bne
p^V 
\sim \frac{E}{2 \kappa (-U)} \qquad \text{as }  U \to 0^-. \ene
From~\eqref{eq:TUUeq}, this causes $T_{UU}$ to diverge at the horizon too: 
\bne T_{UU} (U,V,\phi)  \sim  \frac{E}{r_h \kappa(-U(V)) }  \delta (\phi - \phi_h) \delta (U- U(V)), \qquad U(V) \to 0^-. \label{eq:divst} \ene
From~\eqref{eq:Ucoord}, we see that $T_{UU}$ grows exponentially as $\tW \to -\infty$. This is a blueshift effect. The other components of the stress tensor, such as $T_{VV}$ (which is the same as~\eqref{eq:TUUeq} with a switch $U \leftrightarrow V$), do not diverge at the $U = 0$ horizon. Only one component of the stress tensor diverges in Kruskal coordinates as the particle approaches the horizon, and this is why we changed from global to Kruskal coordinates.

\subsection{Backreaction and the shockwave geometry}
The dominant perturbation to the black hole geometry from the infalling particle is a shockwave due to~\eqref{eq:divst}.
Solving the linearised Einstein's equations for~\eqref{eq:divst} determines the shockwave perturbation $d s^2 \rightarrow d s^2+h_{UU} d U^2$ 
with,~\cite{Fu:2018oaq, Jahnke:2019gxr, Mezei:2019dfv}%
\footnote{There is also a $\prod_{i=2}^{d-1} f(\phi_i)$ factor for the other angular directions, which we have suppressed. It will not affect the final result.}
\bne h_{UU}(U,V,\phi) \sim 16 \pi G_N r_h (-U(V))^{-1}\delta (U- U(V)) f(\phi - \phi_h),\qquad U(V) \to 0^- \label{eq:metrp} \ene
where $f(\phi)$ is the angular profile of the shock (not to be confused with the black hole emblackening factor). When the unperturbed geometry is BTZ, the equation for $f$ is
\bne
r_h^2 f(\phi)- f''(\phi)=\delta(\phi) \label{eq:fphid}
\ene
whose solution is
\bne f(\phi) = \frac{1}{2r_h} \frac{\cosh (r_h(\pi - \tilde \phi))}{\sinh(\pi r_h)} \label{eq:fphi1} \ene 
with $\tilde \phi := \phi \pmod{2\pi}$. The integration constants are fixed by continuity and the jump condition across $\phi =0$. For higher-dimensional AdS-Schwarzschild, $f$ is known~\cite{Shenker:2014cwa} and functionally similar to BTZ's $f$. As can be seen from~\eqref{eq:metrp} and~\eqref{eq:fphi1}, and in all dimensions, the angular profile of the shock is peaked at the position of the particle. $f(\phi)$ in~\eqref{eq:fphi1} has $2\pi$ periodicity and is symmetric about $\phi = \pi$. An equivalent way of writing~\eqref{eq:fphi1}, which is convenient as it is without the mod $2\pi$, is as the Fourier series
\bne f(\phi) = \frac{1}{2\pi} \sum_{n=-\infty}^\infty \frac{e^{in\phi}}{n^2 + r_h^2} .\ene

The metric perturbation~\eqref{eq:metrp} corresponds to a shift in the $V$ coordinate across the shockwave, $V \mapsto V' = V + \Theta (U) \Delta V(\phi)$, with
\bne \begin{split} \Delta V (\phi) \sim \frac{4\pi G_N E}{(-U)} f(\phi-\phi_h) ,\qquad U \to 0^-
\label{eq:delVp}
\end{split} .\ene

\subsection{Correlation function and scrambling time}\label{sec: scrambling time}

Now we will calculate the scrambling time from a bulk computation of a two-sided correlator with a probe operator $V$ (not to be confused with the Kruskal coordinate) in the TFD geometry perturbed by the insertion of the $W$ operator on the right-hand side:%
\footnote{We use the same conventions as~\cite{Shenker:2014cwa} for the definition of left and right boundary operators and their time evolution. 
}
\bne C (\tW) = \braket{\text{TFD} | W_R (\tW,\phiW) V_L (\tV, \phiV) V_R (\tV, \phiV) W_R (\tW,\phiW) | \text{TFD}}. \label{eq:twosi} \ene 
This is a two-point function in the perturbed state $W_R \ket{\text{TFD}}$. The $W$ operators in~\eqref{eq:twosi} are smeared operators, with the smearing finely-tuned so that their insertion into the TFD state is dual to inserting a massive bulk particle near the right AdS boundary at $\tW$ and $\phiW$, and with energy $E$ and angular momentum $J$. We give the details of this smearing kernel in section~\ref{sec: particles}.
$C$ is a function of $(\tW - \tV)$, so we will set $\tV =0$ without loss of generality. We have implicitly set the non-$\phi$ angular positions of $V$ and $W$ to zero, using the rotational symmetry of the problem.
Also, for simplicity, to dimensionally reduce the problem, we assume that the $W$-particle's motion is in the same plane as the $V$ and $W$ insertions, the $(r,\phi)$ plane; that is, we take the non-$\phi$ components of $W$'s angular momentum to vanish.

Note that, while~\eqref{eq:twosi} is not the same as the OTOC~\eqref{eq: OTOC}, they are both analytic continuations to the second sheet of the same Euclidean four-point function~\cite{Shenker:2014cwa}. To be specific, the two-sided correlator is related to the OTOC~\eqref{eq: OTOC} by 
\bne \braket{\text{TFD} | W_R (t) V_L(0) V_R (0) W_R (t) | \text{TFD}} = \langle W(t) V(0)W(t) V(i\beta/2)\rangle_{\beta}.\ene
We work with~\eqref{eq:twosi} for convenience; it is a two-point function straightforwardly computed with the geodesic approximation%
\footnote{This order of limits is necessary so that $V$ does not backreact. There is also an additive constant in this equation that is renormalisation scheme-dependent, though it is fixed given a choice of boundary CFT two-point function normalisation. Its value will not matter in the end, so, for convenience, we set it to zero.}%
\bne \lim_{m_V \to \infty} \lim_{G_N \to 0}\log(C(\tW)) = -m_V L_{\text{ren.}}.
\label{eq:scrit} \ene 
We will see that the CFT calculation of~\eqref{eq: OTOC} and the bulk calculation of~\eqref{eq:twosi} give the same differences in scrambling time, consistent with both correlators probing the same chaotic physics.
 
So, we need the renormalised bulk geodesic distance between the $V$ operator insertions on the left and right boundaries, at $r=\epsilon^{-1}$ and $\phi = \phiV$.
We evaluate this by calculating the length of the $\phi = \phiV$ curve -- the geodesic in the unperturbed geometry -- in the perturbed geometry, which suffices to capture the first-order
correction to $L$ due to the perturbation. If we renormalise $L$ such that $L_{\text{ren.}} = 0$ without the perturbation, then, with the perturbation,
\bne L_{\text{ren.}}  = \Delta V 
+  O(G_N^2)\,. \label{eq:length} \ene
So, since $\Delta V$ grows with decreasing $\tW$, this shows that $C(\tW)$ decreases the earlier that $W$ is inserted with respect to the probe, and thus that the $W$ perturbation destroys the left-right boundary correlation. This is the scrambling effect of the $W$ operator on the TFD state. 

The scrambling time $t_*$ can be defined as the value of $-\tW $ for which the first order $G_N$ correction to $C(\tW)$ becomes leading order, i.e. when $\Delta V$ becomes sufficiently large that the following perturbative expansion breaks down
\bne \log(C(\tW)) = - m_V \Delta V + O (G_N^2) \ene
From~\eqref{eq:delVp}, we see scrambling starts at 
$\kappa t_* = O(\log(G_N^{-1}))$.
This is not a precise definition of the scrambling time, as it only determines the $G_N$ scaling of $t_*$, but we \textit{will} be able to unambiguously define the change in scrambling time $\Delta t_*$. For now, let us pick an arbitrary, small, but $O(G_N^{0})$ constant $a$, and define the scrambling time as when~\eqref{eq:scrit} equals $-a$.
Then, from~\eqref{eq:nhtco},~\eqref{eq:delVp}, and~\eqref{eq:length}, we get
\bne t_{*}( E,J) = \frac{1}{\kappa} \log \left ( \frac{  a  A^{(E,J)}B}{4\pi G_N E \, m_V f(\phiV - \phi^{(E,J)}_h)} \right ). \label{eq:scrmt}\ene
The constant $a$ captures the ambiguity in how $t_*$ is defined. We have added arguments to $A$ and $\phi_h$, to emphasise that these are the quantities that depend on the energy and angular momentum of the $W$ particle. Recall from~\eqref{eq: delta t div} that the coefficient $A^{(E,J)}$ diverges as $J\to J_{\rm crit.}$, and so the scrambling time diverges in this limit, correctly interpolating into the regime in which the particle doesn't approach the horizon.

Now we calculate the difference in scrambling times for two W-particles,  $\Delta t_* := t_*^{(2)} - t_*^{(1)}$, with different conserved energies and angular momenta, keeping everything else fixed, i.e. the unperturbed black hole geometry and the $W$ and $V$ insertion points. From~\eqref{eq:scrmt}, this is
\bne 
\boxed{
\Delta t_* = \frac{1}{\kappa} \log \left (  \frac{E_1 \, A^{(2)} f(\phiV -\phi_h^{(1)} )}{E_2 \, A^{(1)} f(\phiV - \phi_h^{(2)})} \right )
}
\label{eq:delts}\ene
This is valid for both BTZ and AdS$_{d+1}$-Schwarzschild in any dimension, for particles that fall into the black hole. If one of the particles falls in and the other does not, then the difference in scrambling times is infinite.
Eq.~\eqref{eq:delts} is independent of the arbitrary constant $a$, 
so, as promised, $\Delta t_*$ is unambiguous. 
For BTZ, we know what $A$ and $f$ are, and we get
\bne \Delta t_* = \frac{1}{r_h} \log \left (\sqrt{\frac{E_1^2-J_1^2}{E_2^2-J_2^2}} \frac{\cosh(r_h(\pi - \tilde{\varphi} (E_1,J_1)))}{\cosh(r_h(\pi - \tilde{\varphi} (E_2,J_2)))} \right ) \label{eq:delt2}\ene
where, using~\eqref{eq:phihd},
\bne \tilde{\varphi}(E,J) := \left( \phiV -\phiW - \frac{1}{r_h} \arctanh \left( \frac{J}{E}\right ) \right) \mod{2\pi}.
\ene
Eq.~\eqref{eq:delt2} is the difference in scrambling times for global BTZ. The mod $2\pi$ in $\tilde \varphi$ makes it difficult to simplify. 

We will hold off on discussing BTZ's $\Delta t_*$'s features until after taking the planar BTZ limit, because the resulting formulas are simpler while the physics and the qualitative features are unchanged. Unlike AdS-Schwarzschild, taking the planar limit of BTZ does not change the fact that everything falls into the black hole. Taking the planar limit is also necessary for comparison to the CFT results in section~\ref{sec: CFT}.

We take the planar limit by taking $r_h \to \infty$, with $x = \frac{r_h}{R_h} \phi$ the new decompactified coordinate, and $P_x = \frac{R_h}{r_h} J$ the conserved linear momentum, where $R_h$ is the horizon radius in the new coordinates.%
\footnote{To take the planar limit carefully, we start from the BTZ metric in global coordinates, define $x = a \phi$, with $a = r_h/R_h$, and then rescale 
$\tilde t = a t$
and $R=r/a$. Then we can take $r_h \to \infty$ while holding $R_h$ fixed, and the result is the planar BTZ metric in $(\tilde t, R, x)$ coordinates, with $x\in \mathbbm{R}$. In these new coordinates, the horizon is at $R = R_h$, and the temperature is $T = R_h/2\pi$. The momentum is $P_x = J/a$ and the energy with respect to the new $\tilde t$ is $\tilde E = E/a$. Lastly, to not clutter notation, we will drop the tilde from $\tilde E$, though one should keep in mind that $E_{\text{new}}$ is a rescaling of $E_{\text{old}}$, $E_{\text{new}} = \frac{R_h}{r_h} E_{\text{old}}$.
}   
The $W$ perturbation is inserted at $\xW$ and the $V$ probe at $\xV$. 
Taking the $r_h \to \infty$ limit of~\eqref{eq:fphi1} gives
\bne \lim_{r_h \to \infty} r_h f(x) = \frac{1}{2} e^{-R_h |x|} .\ene
This correctly decays as $|x| \to \infty$.

The distance the $W$ particle travels in the $x$ direction, $\Delta x := \lim_{t\to \infty} (x(t) -\xW)$,  equals
\bne \begin{split} \Delta x = \frac{1}{R_h} \arcsinh \left(\frac{P_x}{\sqrt{E^2 - P_x^2}} \right ) = \frac{\eta_x}{R_h}. \end{split} \ene
where $\eta_x$ is the rapidity in the $x$ direction. As usual, the velocity, linear momentum and rapidity are related by $v_x = \tanh \eta_x= \frac{P_x}{E}$.
Note that $\Delta x$ diverges as $P_x \to E$ from below, because then the particle is moving parallel to the AdS boundary and its velocity perpendicular to the AdS boundary is zero. 

Now, let us again give the difference in scrambling time for two $W$-particles, this time for planar BTZ. The conserved energy and momenta of the particles are  $(P_{1,x}, E_1)$ and $(P_{2,x}, E_2)$.
Following the same steps as in the global BTZ case, the difference in the $W$-particle scrambling times is
\bne \boxed{  \Delta t_* = \left |\xW -\xV + \frac{\eta_{2,x}}{R_h}  \right| - \left|\xW-\xV + \frac{\eta_{1,x}}{R_h}\right| + \frac{1}{R_h}\log \left(\frac{E_1 \cosh \eta_{2,x}}{E_2 \cosh \eta_{1,x}} \right)} \label{eq:pldsc} .\ene
This is the planar BTZ simplication of~\eqref{eq:delt2}. We will match this result to a CFT calculation of $\Delta t_*$ in section~\ref{sec: CFT}.

The formulas~\eqref{eq:delts} and~\eqref{eq:pldsc} for the change in scrambling time have two distinct contributions. First, there are, respectively, the $\phi_{\scriptscriptstyle V, W}$ and $x_{\scriptscriptstyle V, W}$-dependent terms, which relate to the angular or $\Delta x$ distance travelled by the particles before reaching the event horizon, and the $\phi_h$ or $x_h$ position at which the particles reach the horizon is the spatial position of the tip of the butterfly cone on the boundary.
The cone then grows with butterfly velocity $v_b=1$, and the time it takes to reach the $V$ probe depends on the probe's relative separation from $\phi_h$ or $x_h$. 
The second contribution to~\eqref{eq:delts} and~\eqref{eq:pldsc}, the $\phi_{\scriptscriptstyle V, W}$ and $x_{\scriptscriptstyle V, W}$-independent terms respectively, come from how long each particle takes to reach a given near-horizon blueshift factor, and this comes from how, for fixed $E$, the particle must start from lower in the AdS potential if $J$ or $P_x$ increase. On the CFT side, there is a corresponding delay in the formation of the butterfly cone. 
We refer to the temporal position of the butterfly cone tip as the global scrambling time, which equals $\min_{\phiV} t_*$ or $\min_{\xV} t_*$, and it is independent of the insertion points of the $W$ and $V$ operators.

We will comment on the features of this formula for
two special cases of $W$-particle kinematics. First, when one $W$-particle is a boosted copy of the other: we take $(E_1,P_{1,x}) = (E,0)$ and $(E_2,P_{2,x}) = (\gamma_x E, \gamma_x E v_x)$. Then $\Delta t_*$ is
\bne \Delta t_* = \left |\xW -\xV + \frac{\eta_{x}}{R_h}  \right| - \left|\xW-\xV \right| .  \label{eq: delta t boost}\ene

\begin{figure}
    \centering
    \includegraphics[width=0.85\linewidth]{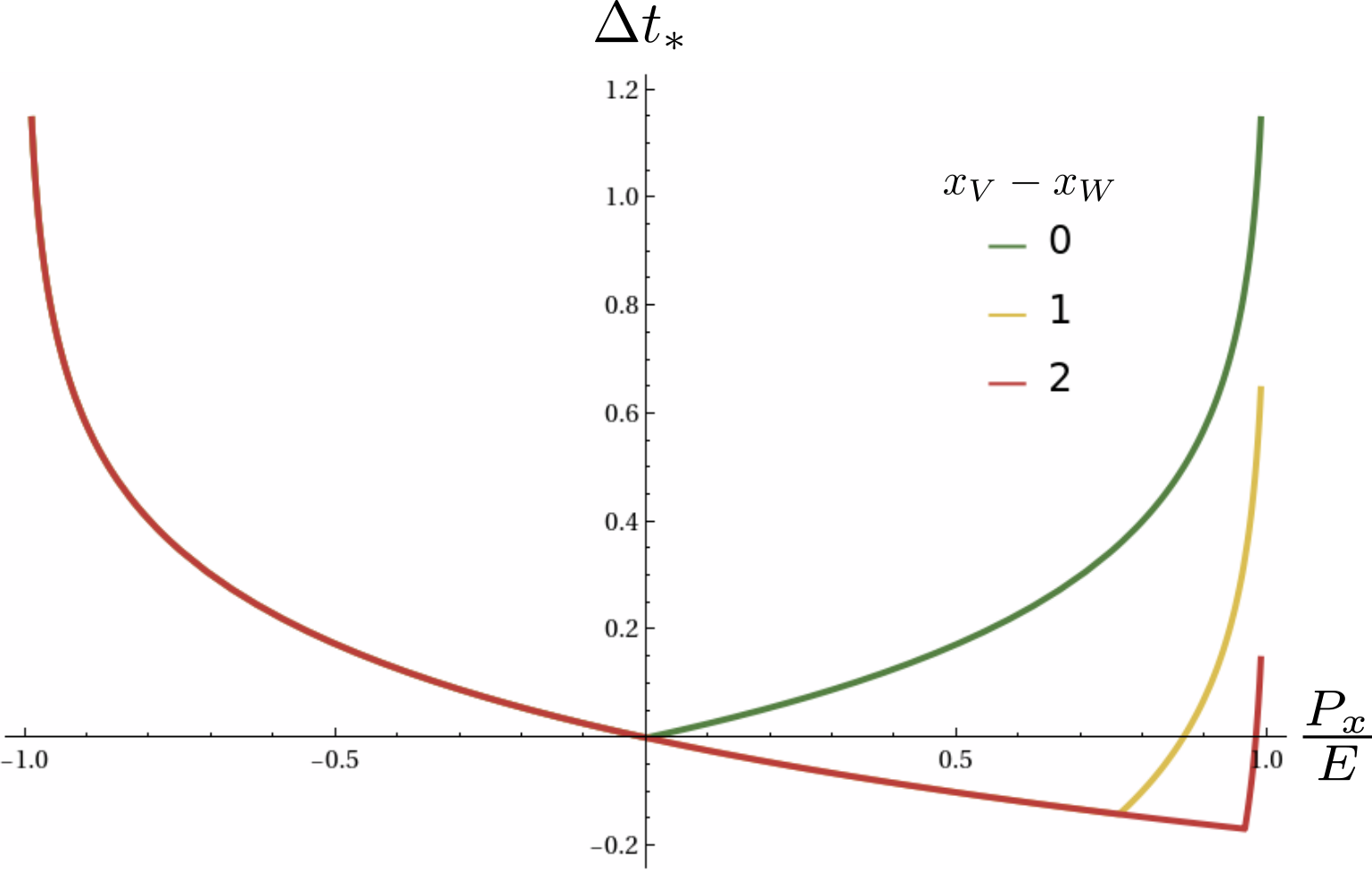}
    \caption{A plot of $\Delta t_*$, Eq.~\eqref{eq:deltat4}, as a function of $P_x / E$ and for different values of $(\xV - \xW)$. The curves overlap for negative values of $P_x/E$.
    }
    \label{fig:delta_t_plot}
\end{figure}

The second kinematic case we consider is where the two $W$-particles have the same energy: 
we take $(E_1,P_{1,x}) = (E,0)$ and $(E_2,P_{2,x}) = (E, P_x)$, with $P_x = E v_x$. Then
\bne  \Delta t_* = \left |\xW -\xV + \frac{\eta_{x}}{R_h}  \right| - \left|\xW-\xV \right| + \frac{1}{R_h}\log \left(\cosh \eta_{x} \right) \label{eq:deltat4} .\ene
As a sanity check, note that $\Delta t_* = 0$ when $v_x =0$, and that, when $\xV = \xW$, $\Delta t_*$ is symmetric under $P_x \to -P_x$.
In Fig.~\ref{fig:delta_t_plot} we plot~\eqref{eq:deltat4} as a function of the $W$-particle velocity $P_x/E = v_x$, for different values of insertion point difference $(\xV - \xW)$.

Below we list some features of the $\Delta t_*$ when the two particles have the same energy, Eq.~\eqref{eq:deltat4}. These features can also be seen in Fig~\ref{fig:delta_t_plot}.
\begin{enumerate}
    \item $\Delta t_* \to +\infty$ as $P_x \to E$ from below. This is because the $W$-particle is moving parallel to the AdS boundary and not any deeper into the bulk. 
    \item When $\xV = \xW$, $\Delta t_* \geq 0$ for any $P_x$. This is both because the second $W$-particle is moving away from the probe's insertion point, and moving more slowly into the bulk.
     \item  
    The scrambling time can decrease if $\xW \neq \xV$. 
    The minimum value of $\Delta t_*$ is at $\eta_x = R_h( \xV -\xW)$, which is when $\xV = x_h$.
    This is as expected: there is a larger effect on the $V$ correlator if we send the $W$ particle towards $\xV$, with a maximal decrease in the scrambling time when the centre of the shockwave intersects the two-sided $V$-probe geodesic.
    \item 
    The change in the scrambling time is independent of $(\xV -\xW)$ if the $W$-particle is moving away from the $V$-probe. Mathematically, this is because
    $|\xV -\xW -\Delta x| - |\xV-\xW| = |\Delta x|$ if $\sgn (\xV -\xW ) \neq \sgn (P_x)$. 
\end{enumerate}

\section{The CFT dual to inserting bulk particles} \label{sec: particles}

In this section, we will determine the CFT operator that creates a bulk particle near the AdS boundary with a particular energy and boundary-parallel conserved momentum.

\subsection{Bulk particle wavepackets and smeared boundary operators.}

\subsubsection{What is a particle?} First, to understand what we mean by bulk particle, we review how to get to classical particle mechanics from QFT, i.e. how and when quantised field excitations behave like point particles following classical trajectories. Readers familiar with this background may wish to proceed to section~\ref{sec:bptbo}, where we construct the smeared CFT operator that creates a bulk particle at a given position and momentum. 

To start, we show how to get to the Schr\"odinger equation from a QFT, i.e. how to take the quantum mechanical limit. Consider the following QFT one-particle amplitude, 
\bne \psi (x,t) = \bra{0} \phi(x,t)^\dagger \ket{\psi}, \label{eq:psieq} \ene
which is the overlap between an arbitrary state $\ket{\psi}$ and
$\phi(x,t)\ket{0}$, which is a one-particle state (in the free approximation).

This will become our quantum mechanical position-space wavefunction. With some caveats, $\psi(x,t)$ can be interpreted as the amplitude for finding a particle for the field $\phi$ at position $x$ in the state $\ket{\psi}$.
The main caveat is that $\phi(x) \ket{0}$ is not really the state of a particle at position $x$, 
in part because there is no position operator in QFTs like there is in quantum mechanics.%
\footnote{One can define a Newton–Wigner operator which reduces to the usual position operator in the non-relativistic limit, $m\to\infty$ or small momentum, but it is neither Lorentz-covariant nor a local QFT operator.}
To be precise, in local relativistic QFTs, there is no observable with support on a localised, bounded region that counts the number of particles in that region, because a particle-counting observable would annihilate the vacuum state, the zero-particle state, and that is not possible because the vacuum is cyclic and separating for local algebras, as follows from the Reeh-Schlieder theorem.

The second caveat 
is that $\phi (x)\ket{0}$ is not in the QFT's Hilbert space because it is not square-normalisable, because of coincident point singularities of local operators, and a rigorous treatment would use operators integrated against test functions~\cite{Haag:1996hvx}. We will not treat this technical detail rigorously.

Next we take the weak coupling limit, as then $\psi(x,t)$ approximately obeys the free field equations of motion, e.g. the Klein-Gordon (KG) equation for a scalar field,
\bne i\hbar \del_t \psi (x,t) = \sqrt{m^2 - \hbar^2 \nabla^2}\, \psi (x,t). \label{eq:KGeqn} \ene
To take the square root, we have implicitly assumed here that $\ket{\psi}$ does not contain negative energy/anti-particle modes, or that they have implicitly been projected out. 
In the non-relativistic limit, for low energy states whose support in the momentum domain satisfies $|p| \ll m$, the Taylor expansion of~\eqref{eq:KGeqn} approximates to the Schr\"odinger equation. 

Next, we show how to get to classical mechanics using the WKB approximation. If we write $\psi(x,t)$ as
\bne \psi (x,t) = |\psi(x,t)| e^{\frac{i}{\hbar}S(x,t)} \label{eq:WKBap} \ene
and plug this into the Schr\"odinger equation, $i\hbar \del_t \psi (x,t) = \hat{H}(x,\nabla) \psi(x,t)$, 
then at leading order in $\hbar$, assuming that $|\psi|$ is slowly varying, we get
\bne -\frac{\del S(x,t)}{\del t} = H(x,\nabla S(x,t)). \label{eq:HJ} \ene
We recognise this as the Hamilton-Jacobi equation, with the amplitude's phase $S$ identified with Hamilton's principal function. Given~\eqref{eq:HJ}, the evolution of the wavepacket's momentum $p(t) = \nabla S(x,t)$ obeys Hamilton's equations of motion, and the centre of a narrow wavepacket will follow the classical trajectory.
This shows how to get to classical mechanics from the non-relativistic, semiclassical limit of QFT.

In the derivation above, we took the non-relativistic limit when we used the Schr\"odinger equation, but note that this is not necessary to reach the classical approximation;
in some cases,
it is possible to derive Lorentz-invariant, relativistic forms of the Hamilton-Jacobi equation~\cite{PhysRev.133.B1622}. 

To give an example that is concrete and similar to what we will consider in the next section, let us take a free scalar field theory and an initial state that is a Gaussian wavepacket
\bne \psi(x,0) = e^{-\frac{(\vec x- \vec x_0)^2}{2\sigma^2}} e^{i \vec p_0 \cdot \vec x /\hbar} . \ene
In the non-relativistic limit, where the dynamics are governed by Schr\"odinger's equation, we can determine the exact solution for $\psi(x,t)$ and find $\langle \hat{\vec x}(t) \rangle = \vec x_0 + \vec p_0 t/m$, 
and $\langle \hat{ \vec p}(t) \rangle = \vec p_0$, showing that the wavepacket follows the classical trajectory, as follows from Ehrenfest's theorem. At $t=0$, $\Delta x (0) = \sigma $, and $\Delta p(0) = \hbar /2\sigma$, saturating the uncertainty principle, and for $t>0$, 
\bne \Delta x (t) = \sigma \sqrt{1+ \left (\frac{\hbar t}{m\sigma^2} \right)^2}, \ene
showing that the particle wavepacket stays localised for $\hbar t/m\sigma^2 \ll 1$. 

\subsubsection{From bulk particle to boundary operator} \label{sec:bptbo}

Following~\cite{Polchinski:1999ry}, we now construct a bulk wavepacket centred on a null geodesic. The WKB approximation requires $r \ll \omega$, because there is a redshift factor in AdS - the frequency at a given radius scales as $1/r$ - and WKB requires the local wavelength to be much smaller than the curvature length scale.
We take $\omega \gg 1$ so that we will be able to match the WKB approximate solution to the large $r$ asymptotic solution. We can use and solve the KG equation in Minkowski spacetime to get an approximate solution, because the spatial width $\omega^{-1/2}$ of the wavepacket is much smaller than the AdS length scale. 
Plugging the WKB ansatz $\phi(x) = A(x) e^{-i\omega f(x)}$ into the KG equation, we get a Gaussian wavepacket solution whose centre is moving in the direction
$\vec e$
\bne \label{eq:bulk_wavepacket} \phi_{\omega, \vec e} (t, \vec x) = e^{-\frac{\omega}{2}(x_\perp^2 + (t - \vec e \cdot \vec x)^2)} e^{-i\omega (t - \vec e \cdot \vec x)} .\ene
This wavepacket has frequency $\omega$ and spatial width $\omega^{-1/2}$, which is sub-AdS scale for sufficiently high frequency. 

To find the boundary operator that creates this bulk wavepacket, we can propagate the solution out to the AdS boundary and use the extrapolate dictionary, i.e., if the bulk field is massless, $\cO (t,\vec x) = \lim_{r\to\infty} r^{d-1} \phi(t,r,\vec x)$.
We determine the envelope function of the large $r$ asymptotic solution by matching the general solution to \eqref{eq:bulk_wavepacket} to the asymptotic large $r$ solution 
in their overlapping regime of validity $1 \ll r \ll \omega$. 

Applying the extrapolate dictionary to the resulting large $r$ form of the wavepacket gives the boundary operator that creates the bulk particle:
\bne \cO_{\omega \vec e}  = \int dt d^{d-1} \hat x K(t+\pi/2, |\hat{x} + \vec e|) \cO (t,\hat x) \label{eq:smeop} \ene
where the kernel $K$ is (up to a constant prefactor)
\bne \label{eq:smea1} \boxed{
K(t,\hat x) = e^{-i\omega t}  e^{- \frac{t^2 + \hat x^2}{\sigma^2}}
}
\ene
and $\hat x$ is the angular direction on the boundary sphere. 

For the bulk wavepacket~\eqref{eq:bulk_wavepacket}, given in~\cite{Polchinski:1999ry}, the spatial width of the boundary smearing function in~\eqref{eq:smea1} is fixed to $\sigma = \sqrt{2/\omega}$. But this choice of $\sigma$ is a special solution to the KG equation
whose transverse spatial profile is constant in the longitudinal direction. We can get an arbitrary $\sigma$ in~\eqref{eq:smea1} with the general solution to the KG equation.

The smeared boundary operator~\eqref{eq:smeop} creates a narrow bulk wavepacket centred on $(r,t,\hat x) = (\omega,-\pi/2,-\vec e)$. 
The radius $r \approx \omega$ is the classical turning point of the bulk wavepacket.
The wavepacket follows the classical trajectory of a particle dropped towards the centre of AdS, which is approximately a null geodesic, even for massive fields, because of starting high in the AdS potential. 

\subsection{How to smear a boundary operator to give bulk particles boundary-parallel momentum}

In this section, we show how to smear a boundary operator so that the dual bulk particle has (conserved) momentum in the direction parallel to the AdS boundary. We will focus on how to give linear momentum, as the CFT will be on a line in the next section.
In this direction, we will calculate the energy and momentum of the state excited from the vacuum by a smeared local operator, $\int K \cO \ket{0}$, to see how they depend on the smearing kernel.

First, we consider a smearing kernel which generates states with zero momentum. In holographic theories, this is the kernel that creates a bulk particle that falls radially in from the boundary, but the discussion here is not limited to holography. 

Suppose that we have a field theory in Minkowski spacetime, and the smearing kernel
\bne \label{eq:smker} K(t,x,\vec y) = e^{-i\omega t} e^{-\frac{t^2+x^2+\vec{y}^2}{\sigma^2} } .\ene
In lightcone coordinates $x^\pm := \frac{1}{\sqrt{2}} (t\pm x)$, the kernel factorises
\bne K(t,x,\vec y) = K_+ (x^+) K_- (x^-) K_y (\vec y) \ene
with
\bne K_\pm (x^\pm) =  e^{-\frac{i \omega x^\pm}{\sqrt{2}}} e^{-\frac{x^{\pm 2}}{\sigma^2}} \label{eq:lcker} \ene
which in the momentum domain is (up to a prefactor)
\bne K_\pm  (p^\pm) = e^{-\frac{\sigma^2}{4}(p^\pm - \frac{\omega}{\sqrt{2}})^2} .\ene
We see that the kernel~\eqref{eq:smker} isolates modes in the smeared operator $\int K \cO$ with $p^\pm = \frac{1}{\sqrt{2}}\omega$ which corresponds to energy $p^t = \frac{1}{\sqrt{2}}(p^+ + p^-) = \omega$ and vanishing linear momentum $p^x = \frac{1}{\sqrt{2}}(p^+ - p^-)$, and $\sigma$ controls the spread in the energy-momentum domain. In App.~\ref{sec:IepsEM}, we calculate the energy and momentum of the $i\epsilon$-regulated state $O(t=i\epsilon)\ket{0}$, which is a complementary way of regulating the state.

We now determine how to smear operators to get non-vanishing momentum. 
To do so,
it suffices to multiply the kernel by $e^{- i p\cdot x}$, but we choose to derive the result from the perspective of boosting the smeared operator. 

Note that the Lorentz transformation of a local scalar operator is
\bne \cO(x) \to \cO'(x) = U(\Lambda)  \cO (x) U(\Lambda)^{-1} = \cO(\Lambda^{-1} x),\ene
so we cannot give a local operator momentum by boosting it.
Instead, consider an arbitrary smearing function $K(y)$ centred on the origin, and the corresponding smeared operator centred at an arbitrary position $x$
\bne\OK (x) = \int d^dy K(y) \cO(y+x) \label{eq:O0def} \ene
If we conjugate this by a boost, then we get
\bne U(\Lambda)  \OK (x) U(\Lambda)^{-1} = \int d^d y K (\Lambda y) \cO (y+\Lambda^{-1} x ) .\ene
This is not quite what we want, because, while the boost changes the momentum of the state $U(\Lambda) \OK (x) U(\Lambda)^{-1} \ket{0}$, boosting $\OK (x)$ also moves the centre of the smeared operator to $\Lambda^{-1}x$.
The smeared operator that we want,  
which inserts an excitation centred at $x$ for a continuous family of boosts,
is $\OKeta (x) := U(\Lambda) \OK(\Lambda x) U(\Lambda)^{-1}$, which is equivalent to
\bne  \OKeta (x) = \int d^dy K(\Lambda y) \cO (y+x) \,.
\label{eq:bstsm} \ene
This is the same as $\OK (x)$ in~\eqref{eq:O0def}, except that the profile of the smearing kernel has been boosted. The kernel $K^\eta$ creates excitations whose energy-momentum is boosted with respect to those created by $K$.

$\OKeta$ is centred on $x$ for all boosts.
Let us check that~\eqref{eq:bstsm} gives us the energy and momentum we expect using the kernel~\eqref{eq:smker}. The boosted smearing kernel, with a boost in the $+ x$ direction, 
is
\bne K^\eta_\pm (x^\pm) = K_\pm (e^{\mp \eta} x^\pm ) \label{eq:boske} \ene
which in the momentum domain is
\bne K^\eta_{\pm} (p^\pm) = e^{\pm \eta} K_\pm (e^{\pm \eta} p^\pm) \ene
and this kernel both changes the widths of the smearing and picks out modes centred on energy and momentum
\bne\label{eq: energy momentum boost} p^t = \omega \cosh \eta, \quad p^x = \omega \sinh \eta .\ene

In the next section, we will use the kernel~\eqref{eq:smea1} and its boosted version~\eqref{eq:boske} to calculate the OTOC of smeared CFT operators dual to the bulk particles considered in section~\ref{sec: bulk}.

\section{CFT$_2$ calculation of the OTOC}\label{sec: CFT}
Consider a 2d Euclidean CFT on the cylinder $\mathbbm{R} \times S^1_\beta$, corresponding to the CFT on a line at finite temperature $T = \beta^{-1}$. In this section, we will calculate the difference between scrambling times, $\Delta t_*$, for two different perturbations of the thermal state, and match to the result calculated for planar BTZ, see eq.~\eqref{eq:pldsc}.

Starting from a Euclidean four-point correlator of two pairs of local scalar operators, $\langle WWVV\rangle_\beta$, we will calculate the following Lorentzian four-point OTOC
\begin{equation}\label{eq: 4 pt smear}
 \GK (\tW, \xW)= \langle \WK (\tW+ i\epsilon_1, \xW) V(i\epsilon_3,0) \WK(\tW+i\epsilon_2,\xW) V(i\epsilon_4,0) \rangle_\beta
 \end{equation}
where $V$ is a local operator. We have set $\tV = \xV = 0$, without loss of generality, and the ordering of the Euclidean times is $\epsilon_1<\epsilon_3<\epsilon_2<\epsilon_4$. If $\WK$ were also a local operator, then this would precisely be the same setup as~\cite{Roberts:2014ifa}, who first calculated the Lyapunov exponent and scrambling time in sparse large-$c$ 2d CFTs. But 
$\WK$ is the smeared operator
\bne \WK(\tW,\xW) = \int dt dx K(t-\tW, x-\xW) W(t,x). \ene 

So, eq.~\eqref{eq: 4 pt smear} is an OTOC of local operators integrated against two kernels: 
\begin{equation}\label{eq: smearing}
    \GK(\tW,\xW) = \int dt_1 dt_2 dx_1 dx_2 K(t_1-\tW,x_1-\xW) K(t_2-\tW,x_2-\tW) \langle W(t_1, x_1) V(0,0) W(t_2,x_2) V(0,0) \rangle_\beta \,.
\end{equation}

For now, $K$ is any test function kernel with some characteristic width and centred around zero. Later, we will take $K$ to be the special smearing kernel~\eqref{eq:smea1} that is finely-tuned such that $\WK$ creates a bulk particle wavepacket with energy $E$ and linear momentum $P_x$, as detailed in section~\ref{sec: particles}. It is precisely this smearing of the local OTOC against kernels that will give us how the scrambling time depends on $E$ and $P_x$, and so allow us to match to the bulk result~\eqref{eq:pldsc}. 

We refer to~\eqref{eq: 4 pt smear} as the smeared OTOC. 
We will calculate the scrambling time from the normalised, connected part of the smeared OTOC~\eqref{eq: 4 pt smear}, which is
\begin{equation}
   \gK (\tW,\xW) = 1-\frac{\GK(\tW,\xW)}{ \langle \WKone \WKtwo \rangle_\beta \langle V_{\scriptscriptstyle 3} V_{\scriptscriptstyle 4} \rangle_\beta} \,.
\label{eq:ncsot}
\end{equation}

\subsection{OTOC of local operators}
In this subsection, we will calculate the OTOC of local operators that we will smear next. This subsection has overlap with older OTOC calculations in, for example,~\cite{Roberts:2014ifa, Perlmutter:2016pkf, Hampapura:2018otw}, but reviewing the derivation, with a few additional details added, makes the section self-contained, orients the reader, and gives us the formulas we will need later.

We start from the Euclidean correlator of four local operators at points on the complex plane, $z_i \in \mathbbm{C}$: 
\bne 1- g(z,\bar z) = \frac{\langle W (z_1,\bar{z}_1) W(z_2,\bar{z}_2) V(z_3,\bar{z}_3)  V(z_4,\bar{z}_4) \rangle}{ \langle W (z_1,\bar{z}_1) W(z_2,\bar{z}_2)\rangle \langle V(z_3,\bar{z}_3)  V(z_4,\bar{z}_4) \rangle} \ene Since this is a Euclidean correlator, $\bar z_i = z_i^*$. The correlator $g$ is a function of the two conformal cross ratios $z$ and $\bar z$, with
\bne z = \frac{z_1-z_2}{z_1-z_3} \frac{z_3-z_4}{z_2-z_4} 
\ene
and $\bar z = z^*$.
To get to a thermal correlator, the conformal map from the plane to the cylinder $\mathbbm{R} \times S^1_\beta$ is
\bne z_i = e^{\frac{2\pi}{\beta} (x_i + i\tau_i)}\ene
with $\tau_i \sim \tau_i + \beta$. The correlator $g(z,\bar z)$  is invariant under this and all conformal maps. The two-point function of a local scalar operator on the cylinder, as a function of the coordinates on the plane, is 
\begin{equation}\label{eq: 2pt local}
     \langle {\cal O}(z_i,\bar{z}_i) {\cal O}(z_j,\bar{z}_j)\rangle_\beta = \left| \frac{2\pi}{\beta} \frac{\sqrt{z_i z_j}}{z_i-z_j} \right|^{2\Delta_{\cal O}}\,,
\end{equation}
In this CFT context, $|(\dots)|^2$ denotes the product of holomorphic and antiholomorphic factors.

Now we analytically continue $g(z, \bar z)$ from the Euclidean section ($\bar z = z^*$) to a Lorentzian correlator. The path $\mathcal{C}$ we take through $\mathbbm{C}^2$ starts on the Euclidean section at $\tau_i = \epsilon_i$, with $\epsilon_i$ infinitesimal, and continues to $\tau_i = \epsilon_i - i t_i$. 
Then we have
\bne z_i = e^{\frac{2\pi}{\beta} (x_i+t_i + i\epsilon_i)}, \quad \bar{z}_i = e^{\frac{2\pi}{\beta} (x_i-t_i-i\epsilon_i)} . \label{eq:zidefs} \ene
The ordering of $\epsilon_i$ determines, and is the same as, the ordering of the operators in the Lorentzian correlator.%
\footnote{Our correlator is a Wightman function, which is an expectation value of products of operators, such as
\bne \bra{0} O_1 (t_1 + i\tau_1) \dots O_n (t_n + i \tau_n) \ket{0}\,. \label{eq:Wight}\ene
With $n$ operators, there are $n!$ Wightman functions. Each Wightman function is a function on $\mathbbm{C}^n$, but their domains are different. For example, the domain of~\eqref{eq:Wight}
is $\tau_1 < \tau_2 < \dots < \tau_n$, because $e^{H \tau_{ij}}$ is only a bounded operator for $\tau_{ij} < 0$. 
} 
On the Lorentzian section, $\bar{z}_i \neq z_i^*$. The analytically continued $g(z,\bar z)$ with $(z,\bar z) \in \mathbbm{C}^2$ has branch points at $z =1$ and $\bar z =1$, the lightcone singularities, 
and following $\mathcal{C}$ can take us around one of these branch points. In appendix~\ref{sec:BP}, we explain how to determine $\Delta \arg\, (z-1)$ as we follow the contour.

Assuming vacuum block dominance, we focus on the Virasoro identity block contribution to $g(z,\bar z)$:
\bne 1- g(z,\bar z) = \mathcal{F}(z) \bar{\mathcal{F}} (\bar z) + \text{non-identity contributions. } \label{eq: id block plus}\ene
Next we take the $c \to \infty$ semiclassical limit,  while keeping $h_v/c$ and $ h_w/c$ fixed, and $h_v/c \ll 1$,
because the identity block $\mathcal{F}$ is known in this regime~\cite{Fitzpatrick:2014vua}:
\bne \mathcal{F} = \left ( \frac{\alpha z (1-z)^\frac{\alpha -1}{2}}{1-(1-z)^{\alpha}} \right)^{2h_v}, \qquad \alpha = \sqrt{1-\frac{24h_w}{c}} . \label{eq:Regli}
\ene
This indeed has a branch point at $z=1$. If we continue around the branch point to the second sheet, then the conformal blocks become, in the $\frac{h_w}{c} \ll z \ll 1 $ regime (the Regge limit)
\bne \mathcal{F}_{II} =  1\pm \frac{24\pi i h_v h_w}{c z} + O(z^{-2})\ene
and 
\bne  \mathcal{\bar{F}}_{II} =  1\pm \frac{24\pi i \bar{h}_v \bar{h}_w}{c \bar z} + O(\bar{z}^{-2}).\ene
The sign is determined by which direction we go around the blocks' respective branch points; for both, going anticlockwise gives the positive sign.

The kernel in our smeared OTOC will localise $t_1$ and $t_2$ around $\tW$, and we are interested in the $-\tW \gg \beta$ regime.
In this limit,
\begin{itemize}
\item  On the principal sheet, $\mathcal{F} \to 1$ and $\bar{\cal{F}} \to 1$.
\item The cross ratios are small. We have
\bne z \approx (z_1 -z_2)(z_4^{-1} -z_3^{-1}), \qquad \bar z \approx (\bar z_3 - \bar z_4)(\bar{z}_2^{-1} - \bar{z}_1^{-1})\,, \label{eq:Regli} \ene
so $z,\bar{z} \approx e^{-\frac{2\pi}{\beta}\min (|t_1|,|t_2|)}$. 
\item Taking the $\epsilon_i$ ordering of~\eqref{eq: 4 pt smear}, using the results in App.~\ref{sec:BP}, we go clockwise the $z = 1$ branch point when $x_2 > x_3$, and clockwise around the $\bar{z} = 1$ branch point $x_2 < x_3$.%
\footnote{Reversing the operator ordering would reverse the direction we go around the branch points.}
So, 
\begin{equation}
\begin{aligned}
    g(z,\bar z) = 
    \begin{cases} \frac{24\pi  h_v h_w}{i\,c z} + O(z^{-2})  \quad {\rm for} \quad x_2 > x_3\\
    \frac{24\pi  \bar{h}_v \bar{h}_w}{i\, c \bar z} + O(\bar{z}^{-2})
    \quad {\rm for} \quad x_3 > x_2.
    \end{cases}
\label{eq:Fitzp}
\end{aligned}
\end{equation}
\end{itemize}

\subsection{Smeared OTOC}

Having derived the OTOC of local operators, eq.~\eqref{eq:Fitzp}, we are in a position to calculate the OTOC with smeared $\WK$,  eq.~\eqref{eq:ncsot}. We set $x_3$, $x_4$, $t_3$ and $t_4$ to zero, and take $W$ and $V$ to be scalar operators. Using~\eqref{eq:Fitzp}, eq.~\eqref{eq:ncsot} becomes 
\begin{equation}
\begin{aligned}
    \gK(\tW, \xW)
    = \frac{6\pi \Delta_v \Delta_w}{i\,c\langle \WK \WK \rangle_\beta }\int \! dt_1 dt_2 dx_1 dx_2&
    K(t_1 - \tW, x_1 - \xW)
    K(t_2 - \tW, x_2 - \xW) \\[4pt]
    &\times
   \langle WW\rangle_\beta 
    \left[\frac{\Theta(x_2)}{z} +
       \frac{\Theta(-x_2) }{\bar z}
    \right].
    \label{eq:gsigma_f}
\end{aligned}
\end{equation}
Next, we take the widths of the kernels in~\eqref{eq: 4 pt smear} to be much smaller than $\beta$. Then the kernel localises $t_1$ and $t_2$ around $\tW$, and $x_1$ and $x_2$ around $\xW$. 
Also, to make further use of the localisation, we change integration variables to the sum and difference of lightcone coordinates $ x_i^\pm = \frac{x_i \pm t_i}{2} $:
\begin{align}
w= x_1^+ + x_2^+, \quad v = x_1^+ - x_2^+, \quad \bar{w} = x_1^- + x_2^-, \quad \bar v = x_1^- - x_2^- .
\end{align}
The $W$ two-point function in these coordinates is
\bne \label{eq:WW2pt}\langle WW \rangle_\beta =\langle W(0,0)W(v,\bar v) \rangle_\beta = \left| \frac{\pi}{\beta} \frac{1}{\sinh(2\pi v/\beta)}\right|^{2\Delta_W}. \ene 

In the narrow kernel limit we have $w \approx \xW + \tW$, $\bar w \approx \xW - \tW$, as well as $v,\bar v \ll \beta$ which gives us the cross ratio approximations
\bne z \approx -\frac{4\pi}{\beta} \epsilon_{34} e^{\frac{2\pi}{\beta}w}v, \quad \bar z \approx \frac{4\pi}{\beta}\epsilon_{34} e^{-\frac{2\pi}{\beta}\bar w}\bar v\ene
where $\epsilon_{34} := e^{-\frac{2\pi }{\beta}i\epsilon_3}-e^{-\frac{2\pi }{\beta}i \epsilon_4}$.
Using these approximations, and performing a $\xW$ and $\tW$ shift in the integration variables,~\eqref{eq:gsigma_f} becomes
\bne \begin{split}\gK(\tW,\xW) \approx \frac{3 i \beta \Delta_v \Delta_w e^{-\tW }}{2c\, \epsilon_{34} \langle \WK \WK \rangle_\beta }\int &K(t_1, x_1)\,
    K(t_2, x_2) \langle WW\rangle_\beta \\
    &\times  \left(\frac{e^{-\frac{2\pi}{\beta}(w+\xW)}}{v} \Theta (x_2 + \xW)- \frac{e^{\frac{2\pi}{\beta}(\bar w +\xW)}}{\bar v} \Theta (-x_2-\xW) \right)    \,.
    \label{eq:simpgK}
\end{split}\ene

Next, we take $|\xW|$ to be larger than the kernel width, so that the overlap the $x_{2,3}$-tails of the smeared $\WK$ with $V$ in the OTOC~\eqref{eq:gsigma_f} is negligible; then, using also that $x_2 \approx \xW$, we have $\Theta (\pm ( x_2+\xW)) \approx \Theta (\pm \xW)$, which simplifies the integral. Following that, assuming that $K$ is an even function of $x_i$, we do the substitution $x_1 \to -x_1$ and $x_2 \to -x_2$ for the second term in~\eqref{eq:simpgK}. This maps $\bar v \to -v$ and $\bar w \to -w$, and does not change $\langle WW \rangle_\beta$. Then~\eqref{eq:simpgK} simplifies further to
\begin{equation}
    \gK(\tW,\xW) \approx  \frac{3 i \beta \Delta_v \Delta_w \IK  }{ 2c\,\epsilon_{34} \, \langle \WK \WK\rangle_\beta }   e^{-\frac{2\pi}{\beta}( \tW + |\xW|)}  \,.
    \label{eq:finalgK}
\end{equation}
where $\IK$ is the $K$-dependent constant
\begin{equation}\label{eq: I integral}
    \IK := \int d w\, dv\, d\bar w \,d\bar v\, \langle WW \rangle_\beta  K(t_1,x_1)K(t_2,x_2) \frac{e^{-\frac{2\pi}{\beta} w }}{v}\,.
\end{equation}
We have 
left the $i \epsilon_i$ implicit in this expression.

Eq.~\eqref{eq:finalgK} tells us that the Lyapunov exponent is $\lambda_L=\frac{2\pi}{\beta}$, because $\IK$ is independent of $\tW$, and that the butterfly velocity is $v_b=1$, because $\IK$ is also independent of $\xW$.
Furthermore, at leading order in $c$, the scrambling time $t_* = \frac{2\pi}{\beta} \log c + O (c^0)$, and this is unaffected by the kernel. The choice of $K$ \textit{will} affect the leading order result for $\Delta t_*$, the difference in scrambling times for two different kernels $K^{(1)}$ and $K^{(2)}$, as well as the center of the butterfly cone. The leading-order results for $\lambda_L$, $v_B$ and $t_*$ are not new, but one thing that is new is that we have shown that these quantities are unaffected by the choice of kernel $K$, with the assumptions and approximations we have made. In our narrow kernel approximation, the smeared operators are still approximately local with respect to the thermal scale. We expect that including subleading corrections in the kernel width would blur the edge of the butterfly cone.

\subsection{Boosted operators}

Before providing explicit results for a given smearing kernel $K$, let us first consider how the OTOC changes when the $W$ operators are smeared with $K^\eta$
\begin{equation}
    \WKeta (\tW,\xW) = \int dx dt K^\eta(t,x) W(t+\tW,x+\xW)\,,
\end{equation}
where $K^\eta$ is the smearing kernel corresponding to the boosted excitation, 
given by~\eqref{eq:boske}.
The smeared two-point functions using $K^\eta$ and $K$ are related by
\bne \langle \WKeta \WKeta \rangle_\beta = \langle \WK \WK \rangle_{\beta_L, \beta_R} \ene
with $\beta_{L,R} = e^{\pm \eta} \beta$. 
In our narrow kernel approximation, 
\bne \langle \WKeta \WKeta \rangle_\beta \approx \langle \WK \WK \rangle_\beta, \ene
because $\langle W(0) W(x) \rangle_\beta$ is approximately $\langle W(0)W(x) \rangle_{\rm{vac.}}$ for $x \ll \beta$, and the vacuum two-point function is Lorentz invariant. We also have the relation
\begin{equation}
\begin{aligned}
    \IKeta (\beta) = e^{\eta} \IK(\beta e^{\eta})\,.\label{eq:IKtransform}
\end{aligned}
\end{equation}
The $e^\eta$ prefactor comes from the boost symmetry-breaking factor of $1/v$ in $\IK$. Note that the $v$ and $\bar{v}$ in~\eqref{eq:simpgK} transform under boosts with opposite signs of $e^{\pm \eta}$. 

All together, we find
\bne \gKeta(\tW,\xW) \approx  \frac{3 i \beta \Delta_v \Delta_w \IK(\beta e^{- \aleph \eta})  }{ 2c\, \epsilon_{34}  \langle \WK \WK\rangle_\beta }   e^{-\frac{2\pi}{\beta}( \tW + |\xW+ \frac{\beta}{2\pi} \eta|)}  \,. \ene
where $\aleph = \sgn (\xW+\frac{\beta}{2\pi} \eta)$. 
Compared to the unboosted kernel results, the main difference is that the center of the butterfly cone is shifted to $\xW+ \frac{\beta}{2\pi} \eta$.
As before, the finite shift of the scrambling time depends on the smearing kernel. We will now compute this finite shift for the smearing kernel~\eqref{eq:smker} which produces a localised particle excitation in the bulk.

\subsection{Specialising to the particle-creating kernel}

We can now compute $\IK $ and $\langle \WK \WK \rangle_\beta $ for the case in which the smearing kernel is given by~\eqref{eq:smker}. We normalise the kernel to $\int K = 1$, though the choice of normalisation does not affect $\Delta t_*$. Because we took the narrow kernel approximation, which made $\langle \WK \WK \rangle_\beta$ boost-invariant, its value not affect $\Delta t_*$, but we give it here for completeness:%
\begin{equation}
\begin{aligned}
    \langle \WK \WK \rangle_\beta  & \approx \frac{e^{\frac{\sigma^2 \omega^2}{2}}}{\pi^2 \sigma^4} \left| \left(\frac{1}{2}\right)^{\Delta_w} \int dw dv e^{-\frac{w^2 + v^2 }{\sigma^2} + i \omega w}  v^{-\Delta_w} \right|^2\\
   &=  \frac{\pi}{  (2\sigma)^{2\Delta_w}\Gamma\left(\frac{1+\Delta_w}{2}\right)^2 } \,.
\end{aligned}
\end{equation}

The $\IK $ appearing in the smeared four-point function is 
\begin{equation}
\begin{aligned}
   \IK  (\beta) &\approx  \frac{e^{\frac{\sigma^2 \omega^2}{2}}}{\pi^2 \sigma^4} \int \left|  dw dv \frac{ e^{-\frac{ w^2 +v^2}{\sigma^2} + i \omega w}}{{(2v)}^{\Delta_w}} \right|^2 \frac{e^{\frac{2\pi}{\beta} w}}{v}\\ 
   & = \frac{i\sqrt{\pi}}{2^{\Delta_w} \sigma^{2\Delta_w +1} \Gamma(1+\Delta)} \exp \left(\frac{\pi  \sigma ^2 (\pi +i \beta  \omega
   )}{\beta ^2} \right)
\end{aligned}
\end{equation}

Together, these give 
\bne \gKeta(\tW,\xW) \approx  -\frac{3\beta \Delta_v \Delta_w 2^{\Delta_w-1}}{c\, \epsilon_{34}\sigma \sqrt{\pi}} \frac{\Gamma(\frac{1+\Delta_w}{2})^2}{\Gamma(1+\Delta_w)}\exp \left(\frac{\pi  \sigma ^2 (\pi +i \beta'  \omega
   )}{\beta'^2} \right) e^{-\frac{2\pi}{\beta}( \tW + |\xW+\frac{\beta}{2\pi}\eta|)}  \,. \label{eq:gketaf} \ene
where $\beta' = \exp(-\sgn(\xW + \frac{\beta}{2\pi})\eta)\beta$. 
The formula for $\gK$ follows from setting $\eta =0$ in~\eqref{eq:gketaf}.
The shift in $\xW$ comes from the $e^\eta$ prefactor in~\eqref{eq:IKtransform}.

The first onset of scrambling, the time at which the tip of the butterfly cone forms, $\min_{\xW}  t_*$, is unaffected by the value of $\eta$. 
However, there is a change in scrambling time that is purely due to kinematics: the tip of the butterfly cone for the OTOC~\eqref{eq:gketaf} shifts by $\frac{\beta}{2\pi} \eta$, and the edge of the cone travels ballistically with butterfly velocity $v_b=1$. 
Correspondingly, depending on the location at which we probe the state, the time required to measure the perturbation will change.
The corresponding change in the scrambling time is
\begin{equation}\label{eq: delta t CFT}
    \Delta t_* =|\xW| - \left|\xW+\frac{\beta}{2\pi} \eta \right|\,,
\end{equation} 
This $\Delta t_*$ agrees with the bulk computation of the same, Eq.~\eqref{eq: delta t boost}. 

As in the bulk computation, we can also compare two particles of arbitrary energies and momenta. The energy scale of $\WK$ is $E \propto (\sigma^+ + \sigma^-)^{-1}$, with $\sigma^+ = \sigma^-$ when $\eta = 0$.
From how $\sigma_\pm$ transforms under boosts, see eqn.~\eqref{eq:boske} and App.~\ref{sec:IepsEM}, this energy scale transforms under boosts to $E \cosh\eta$. The result for $\Delta t_*$ is the same as the bulk calculation, eq.~\eqref{eq:pldsc}.

We have worked in the small kernel width limit, which is similar to taking the point-particle approximation of the bulk wavepacket in section~\ref{sec: bulk}. 
We expect that subleading terms in the kernel width 
would smoothen out the kinks in the function $\Delta t_*$ that can be seen in Fig.~\ref{fig:delta_t_plot}.

\section{Discussion} \label{sec:discussion}

In this paper, we have derived new results for the scrambling behaviour of excitations in holographic CFTs. First, we performed a bulk computation of the scrambling time for BTZ and AdS Schwarzschild black holes and their dependence on conserved energy and momenta. Our main results here are the differences in scrambling times (whose dependence on energy and momenta is leading order in $N$) given by~\eqref{eq:delts} and~\eqref{eq:pldsc}. As a function of particle angular momentum $J$, the scrambling time increases as $J$ increases, up to $J_{\text{crit.}}$, given in Eq.~\eqref{eq:Jcrit}, at which point it diverges. Next, to set ourselves up for a CFT computation of the same results, we derived how to smear a local CFT operator such that it excites a bulk particle with the desired energy and momenta. Lastly, we performed the CFT computation of the OTOC on the thermal cylinder and matched it to our bulk scrambling time results for planar BTZ.

Our work was inspired by considering infalling versus bound radially-oscillating particle geodesics in AdS black hole geometries, and the implication that there are dual operators that do not thermalise but instead oscillate in size.
In vacuum AdS, a particle released from the boundary will also oscillate back and forth, but this is not a puzzle from the CFT perspective because the state is a superposition of a single-trace primary and its descendants, whose energy levels are evenly spaced, so short-time revivals of the state happen. In contrast, if we perturb a black hole state $\cO_H \ket{0}$ with our ``W-particle" operator $W_K$, the OPE will include multi-trace operators. In the strict large $N$ limit, the bulk theory is free and the dimensions of these multi-trace operators are additive, again leading to short-time revivals. At finite $N$, bulk interactions give anomalous dimensions to the multitrace operators~\cite{Liu:1998th}, which one expects to make the frequencies in $\WK \cO_H \ket{0}$ incommensurate, leading to dephasing and thermalisation. But the bulk has both quasinormal and (approximately) normal modes, corresponding to infalling and oscillating orbits respectively, and this suggests that the finite temperature CFT has both a high-$J$ quasi-integrable and low-$J$ chaotic sector. The late-time fate of a perturbation depends on its support in these sectors. Similar behaviour has been studied in, for example,~\cite{Festuccia:2008zx, Hashimoto:2023buz}. 

We did not consider CFTs on $\mathbbm{T}^2$, dual to, at high temperatures, the global BTZ black hole. For CFTs on $\mathbbm{R} \times S_\beta^1$, through its conformal equivalence to the plane, the semiclassical Virasoro blocks are known and are broadly speaking insensitive to the properties and kinematics of the perturbing operator; all perturbations scramble. This is consistent with the bulk side; no particle can avoid falling into a planar BTZ black hole. The same is true for global BTZ, so, on the boundary side, one would expect the same perturbation-insensitivity of the OTOC for a CFT on $\mathbbm{T}^2$. But, unlike the cylinder, the torus is not conformally flat, and there are no closed expressions for torus Virasoro blocks that can be continued to the OTOC configuration like we did in section~\ref{sec: CFT}.

We only did the CFT computation for $d=2$. In higher dimensions, there is richer behaviour on the bulk side. In particular, the absence of chaotic dynamics for perturbations above $J = J_{\text{crit.}}$. But, as for the torus, there are difficulties in calculating OTOCs on the boundary side for higher dimensions.
Firstly, while for AdS$_3$/CFT$_2$ all bulk graviton exchanges are resummed and contained in the Virasoro identity block, in higher dimensions, the equivalent would be to resum over all the stress tensor and multi-stress tensor block contributions; not an easy task, though see~\cite{Huang:2024wbq, Haehl:2025ehf} for progress in this direction. 
Secondly, even if the manifold is locally conformally flat, there are global obstructions to conformally mapping $M^{d-1} \times S_\beta^1 $ to $\mathbbm{R}^d$. One exception, where it is possible to calculate a thermal OTOC (using an EFT approach rather than attempting to resum the blocks), is for a CFT on $\mathbbm{H}^{d-1}\times S^1_\beta$, because of its conformal equivalence to the Rindler wedge when $\beta = 2\pi$~\cite{Haehl:2019eae}. 
But such CFTs are dual to topological black holes with hyperbolic horizons~\cite{Casini:2011kv}, and, just as for BTZ black holes, no massive or massless particle can avoid falling through the horizon; therefore, we cannot investigate the transition to non-scrambling behaviour in this setup.

We focused on non-rotating BTZ and AdS black holes, but OTOCs have also been calculated for rotating black holes~\cite{Poojary:2018esz, Jahnke:2019gxr, Mezei:2019dfv, Blake:2021hjj, Craps:2021bmz}. The rotation leads to a splitting of the Lyapunov exponent into non-equal left and right Lyapunov exponents~\cite{Jahnke:2019gxr},
and an oscillatory modulation of the OTOC decay~\cite{Mezei:2019dfv}. A particle with angular momentum in a static black hole background is physically distinct from a particle without angular momentum in a rotating black hole background. It would be interesting to explore the interplay between the black hole's angular momentum and the particle's angular momentum.

In the bulk, we have worked in the large-$N$ semi-classical limit. 
We have approximated the particle wavepacket as a classical point particle, and so missed some finite $N$ effects. 
For example, as we have discussed, when $J > J_{\text{crit.}}$, the classical particle will not reach the horizon, but at finite $N$, a fraction of the particle wavepacket will tunnel through the angular momentum barrier each time it bounces off of it, giving a small imaginary part to the boundary quasiparticle's frequency~\cite{Festuccia:2008zx}. Through this channel, the excitation \textit{will} eventually scramble, though at a rate that is exponentially suppressed in $N$, $\Gamma \approx e^{-N^2J (\cdots)}$. 
 Also, the particle will emit gravitational radiation as it orbits, losing angular momentum and energy and eventually falling in, and this is perturbatively suppressed in $1/N$. Both of these effects, and the similarity to many-body scars which our non-thermalising states share, were considered in~\cite{Dodelson:2022eiz}. At finite $N$, there is also a delocalisation timescale for the wavepacket, when the point particle approximation breaks down. All these time scales can be made parametrically longer than the AdS time scale. Lastly, besides finite $N$, there are also finite $\lambda$ stringy corrections to scrambling that one could consider in our context~\cite{Shenker:2014cwa}.

Besides those phenomena that we have already discussed, there are other predictions from the bulk that are curious from the boundary perspective. For example, suppose we send two bulk particles from the boundary of AdS-Schwarzschild with oppositely oriented angular momenta. With a fine-tuning of the kinematics, these particles can orbit the black hole an arbitrary number of times before colliding and falling into the black hole.  
On the boundary side, this will look like a pair of excitations travelling around the sphere, oscillating in size, refusing to thermalise, sometimes even passing through each other. Only when the bulk particles are at the same angular and radial depth can they collide and fall in, and then the boundary excitations thermalise and scramble, and this is \textit{highly} sensitive to the fine-tuned kinematics.

\acknowledgments

We would like to thank Ben Craps, Marius Gebershagen, Felix Haehl, Henry Maxfield, Mark Mezei, Andrei Parnachev, Christoph Uhlemann, and Mark van Raamsdonk for useful discussions, and Dong Ming He and Maria Kynsh for initial collaboration. Work at VUB is supported by FWO-Vlaanderen project G012222N, and by the Vrije Universiteit Brussel through the Strategic Research Program High-Energy Physics. The work of JH is also supported by FWO-Vlaanderen through a Junior Postdoctoral Fellowship 12E8423N, and by Taighde Eireann – Research Ireland under Grant number SFI-22/FFP-P/11444. The work of AR is also supported by FWO-Vlaanderen through a Senior Postdoctoral Fellowship 1223125N. 

\appendix

\section{Energy and momentum of an \texorpdfstring{\boldmath$i\epsilon$}{i\epsilon}-regulated local operator insertion.} \label{sec:IepsEM}
\subsection{Imaginary time}
Consider an unnormalised state which is the vacuum excited by a local operator insertion in imaginary time:
\bne \ket{\psi_\epsilon} = \cO(t=i\epsilon)\ket{0} = e^{-H\epsilon} \cO(0) e^{H\epsilon}\ket{0} \label{eq:iepss}\ene
Using 
\bne \frac{d}{d\epsilon} \cO(\pm i\epsilon) = \mp [H,\cO(\pm i\epsilon)] \ene
we have
\bne \frac{d}{d\epsilon} \braket{\psi_\epsilon | \psi_\epsilon} = -2 \bra{\psi_\epsilon}H \ket{\psi_\epsilon}\ene
so
\bne E = \frac{\bra{\psi_\epsilon}H \ket{\psi_\epsilon}}{\braket{\psi_\epsilon | \psi_\epsilon}} = -\frac{1}{2}\frac{d}{d\epsilon} \log \braket{\psi_\epsilon | \psi_\epsilon} .\ene
$\braket{\psi_\epsilon | \psi_\epsilon}$ is a two-point Wightman function: $\bra{0}\cO(-i\epsilon) \cO(i\epsilon) \ket{0}$.

Now, if we assume that the theory is conformal and that $\cO$ is a scalar primary operator, then $\braket{\psi_\epsilon | \psi_\epsilon} = (2\epsilon)^{-2\Delta}$ and $E = \frac{\Delta}{\epsilon}$.

If we changed the state to $\cO(t+i\epsilon)\ket{0}$, giving the operator some Lorentzian time, we would get the same result.

\subsection{Complex time and space}
Now we generalise further. Take the state
\bne \ket{\psi} = \cO(x) \ket{0} \ene
with $x$ complex:
\bne \cO(x^\mu) = \cO(x_R^\mu + i x_I^\mu) = e^{P \cdot x_I} \cO(x_R) e^{-P\cdot x_I}.\ene
$\cO(x_R+ix_I)^\dagger = \cO(x_R-ix_I) = \cO(x^*)$ so $\bra{\psi} = \bra{0} \cO(x^*)$. 
For non-real $x$, $\braket{\psi|\psi} \neq 0$.

Using $\del_{x_I^\mu} \cO = -[P_\mu, \cO]$ and $\del_{x_I^\mu} \cO^\dagger = [P_\mu, \cO^\dagger]$, we have
\bne \del_\mu \bra{0} \cO(x^*) \cO(x) \ket{0} = -2 \bra{0} \cO(x^*) P_\mu \cO(x) \ket{0} \ene
and so
\bne \langle P_\mu \rangle = -\frac{1}{2} \del_{x_I^\mu} \log \bra{0} \cO(x^*) \cO(x) \ket{0}. \ene
For a conformal theory, using the conformal 2-point function, this becomes
\bne \langle P_\mu \rangle_\psi = \Delta \frac{x_{I,\mu}}{|x_I|^2} \ene
With this, we can calculate the energy and momentum of the state excited by the boosted operator
\bne \cO (t+i\epsilon \cosh \eta, x + i\epsilon \sinh \eta) \ene
and get
\bne \langle P_0 \rangle = \frac{\Delta \cosh\eta}{\epsilon} , \quad \langle P_1 \rangle= \frac{\Delta \sinh \eta}{\epsilon} .\ene

\subsection{Finite temperature}
Now we're interested in the energy and momentum of the perturbed thermal density matrix
\bne \rho = \frac{\cO(x) \rho_\beta \cO(x)^\dagger}{\Tr (\cO(x) \rho_\beta \cO(x)^\dagger)} \ene
where $x^\mu$ can be complex.

We find
\bne \Tr (\rho P_\mu) = -\frac{1}{2} \del_{x_I^\mu} \log \Tr (\cO \rho_\beta \cO^\dagger) + \frac{\Tr (P_\mu \rho_\beta \cO^\dagger \cO)}{\Tr(\rho_\beta \cO^\dagger \cO)}. \ene

For $|x_I| \ll 1$, the energy and momentum of excitation dominate over that of the thermal background. Indeed, using that $\cO^\dagger \cO = \frac{\mathbbm{1}}{|2x_I|^{2\Delta}}$ plus less singular terms, we have
\bne \Tr (\rho P_\mu) \sim \Delta \frac{x_{I,\mu}}{|x_I|^2} + \Tr(\rho_\beta P_\mu), \qquad |x_I| \to 0. \ene
This is the same as the vacuum result, with a correction from the $\langle P_\mu \rangle_\beta$ of the thermal background.

\section{Branch point analysis: the change in \texorpdfstring{\boldmath$\arg\,(z-1)$}{arg(z-1)}} \label{sec:BP}
Consider the conformal cross ratio
\bne z-1 = - \frac{z_{14}z_{23}}{z_{13}z_{24}} .\ene
Supposing $z_i$ are functions of a time parameter $t$, we want to know how the dependence of the argument of $z-1$ on $t$. If $\arg \,(z-1)$ changes by $2\pi$ as $t$ increases, then we have gone once anticlockwise around the $z=1$ branch point.

First, we use that
\bne \arg\left( z-1 \right) = -\pi + \arg(z_{14}) + \arg (z_{23}) - \arg (z_{13}) - \arg(z_{24})\ene
where
\bne \label{eq:argz} \arg (z_{ij} ) = \arctan \left (  \frac{|z_i| \sin \epsilon_i - |z_j| \sin \epsilon_j }{|z_i| \cos \epsilon_i - |z_j| \cos \epsilon_j}\right), \qquad z_i = |z_i| e^{i\epsilon_i} .\ene
Next, we note that $\arctan(x)$ jumps by $\pi$ when the denominator of its argument passes through zero:
\bne \left[\arctan \left(\frac{a(x)}{x-x_0}\right)\right]^{x=x_0+0^+}_{x=x_0-0^+} = \text{sign}(a(x_0)) \pi\ene
Away from the zero of the denominator, $\arg(z_{ij}) = O(\epsilon)$, so the only contribution to $\Delta (z-1)$ is from the jumps.
The zero of our denominator in~\eqref{eq:argz} is at
\bne |z_i| \cos \epsilon_i = |z_j| \cos \epsilon_j \ene
and, at this point, the sign of our numerator is
\bne \text{sgn} (|z_i| \sin \epsilon_i - |z_j| \sin \epsilon_j) = \text{sgn} (\epsilon_i - \epsilon_j) .\ene
If $\arctan(x)$ jumps from $+\pi$ to $-\pi$, it's because the angle has wound around the anticlockwise direction, so the jump in $\arctan(x)$ is minus $\Delta \arg(z_{ij})$.
Therefore, the change in the argument from the jumps is
\bne \Delta \arg (z-1) = \pi (c_{13}+c_{24}-c_{23}- c_{14} ) \ene
where
\bne 
c_{ij} := \text{sgn}(\epsilon_i - \epsilon_j) [\Theta (|z_i (t)|  -|z_j (t)| )]^{t_{\rm{final}}}_{t_{\rm{initial}}} .\ene

Now we apply this result to our setup. When $\epsilon_1 < \epsilon_3 < \epsilon_2 < \epsilon_4$, and $|z_3| = |z_4| = 1$, $|z_1| = e^{\tW+x_1}$, and $|z_2| = e^{\tW+x_2}$, then, for the path from $\tW=0$ to $\tW= -\infty$,
\bne \Delta \arg(z-1) = 2\pi \Theta (x_2), \ene
i.e. it winds anticlockwise around the branch point.

For $\arg(\bar z -1)$, we see from~\eqref{eq:zidefs} that both the direction of time and the $\epsilon_i$ ordering are effectively reversed. The result is
\bne \Delta \arg(\bar z-1) = 2\pi \Theta (-x_2). \ene
\bibliographystyle{JHEP}
\bibliography{references.bib}

\end{document}